\documentclass{article}
\usepackage{spconf,amsmath,graphicx, amsfonts,hyperref}
\title{Voice-preserving Zero-shot Multiple Accent Conversion}
\name{Mumin Jin$^{1,2}$, Prashant Serai$^{1}$, Jilong Wu$^{1}$, Andros Tjandra$^{1}$, Vimal Manohar$^{1}$, Qing He$^{1}$ \thanks{Work was done when Mumin was an intern at Meta AI}\thanks{Correspondence to Prashant Serai: pserai@meta.com}}
\address{$^1$Meta AI, $^2$ MIT}
\begin{document}
%
\maketitle
\begin{abstract}
Most people who have tried to learn a foreign language would have experienced difficulties understanding or speaking with a native speaker's accent. For native speakers, understanding or speaking a new accent is likewise a difficult task. An accent conversion system that changes a speaker’s accent but preserves that speaker’s voice identity, such as timbre and pitch, has the potential for a range of applications, such as communication, language learning, and entertainment. Existing accent conversion models tend to change the speaker identity and accent at the same time. Here, we use adversarial learning to disentangle accent dependent features while retaining other acoustic characteristics. What sets our work apart from existing accent conversion models is the capability to convert an unseen speaker’s utterance to multiple accents while preserving its original voice identity. Subjective evaluations show that our model generates audio that sound closer to the target accent and like the original speaker.

\end{abstract}
\section{Introduction}
\label{sec:intro}
For many people, their own or others' accents present a severe obstacle  to communication. For some other people, watching a movie set in America with British accented speakers creates a dissonance that takes them out of the immersive experience. Therefore, a system that converts a speaker's accent yet still preserves the original voice identity can have a great impact in a wide range of situations including communication, language learning, and entertainment. 

The main challenge in voice-preserving accent conversion is the need to disentangle features related to a speaker’s voice, accent, and linguistic contents. Usually, each speaker in a data set only speaks with one accent, and there is relative scarcity in the available number of speakers and audio recordings for non-native accents. Previous works have used adversarial learning to disentangle features \cite{multilingual_tts, tianjin_mandarin}, where both apply a discriminator to wipe-out speaker dependent information from content embeddings. Other works, such as \cite{style_content_disentanglement, discrete_disentanglement}, achieve disentanglement via quantization of different features to obscure undesired information. 

Conventional accent conversion approaches require the availability of reference utterances with the same text with target accents, during synthesis \cite{accentron, PPG_ac, ppg_frame_pairing, vc_for_accent_reduction}. The applications of these approaches are very limited, as we often do not have access to reference utterances with the same linguistic content in a different accent. Recently, \cite{key_baseline, referencefreeFAC, newzeroshotpaper} have proposed systems that convert accents without needing a reference utterance during inference. The systems proposed in both \cite{key_baseline} and \cite{referencefreeFAC}, however, are not zero-shot accent conversion systems because they require further training on the input utterances. In the case of \cite{key_baseline}, the ASR component needed to be fine-tuned with the input speaker's utterances, and in the case of \cite{referencefreeFAC}, a dedicated model for each new speaker must be trained on parallel speech. The authors of \cite{key_baseline} also acknowledge that their converted utterances were perceived to have a different voice from that of the original utterance. Our proposed system differs from the zero-shot, reference-free accent conversion system from \cite{newzeroshotpaper} in that our system allows for synchronous accent conversion, and to multiple native and non-native accents. 

Our proposed system is most similar to the accent conversion models proposed in \cite{key_baseline}, \cite{newzeroshotpaper}, and \cite{tianjin_mandarin}. Unlike these works, our model converts unseen utterances with arbitrary accent to utterances with multiple target accents. Listeners from our perceptual tests agree that our model is good at preserving the original speaker's voice characteristics. In our model, we utilize a pre-trained model checkpoint to extract speaker and accent independent text predictions prior to training. We further disentangle the accent-dependent features from other features with an accent discriminator. Finally, the processed, disentangled features are re-combined and fed to a HiFiGAN decoder to reconstruct the audio waveform. Our work make contributions in three major ways: 1) To the best of our knowledge, our model is the first to convert arbitrary accented, unseen speech to multiple target accents while preserving non-accent related voice characteristics. 2) We do not require text labels associated with accented speech or speaker ID labels during training, although we use an existing ASR model checkpoint trained on native accented English speech to extract linguistic features. 3) We convert accent while keeping the output synchronized to input, allowing for applications such as dubbing a video with different accents. 

\section{Proposed System}
\label{sec:proposed}
\begin{figure}[htb]

\begin{minipage}[b]{1.0\linewidth}
  \centering
  \centerline{\includegraphics[width=8.5cm]{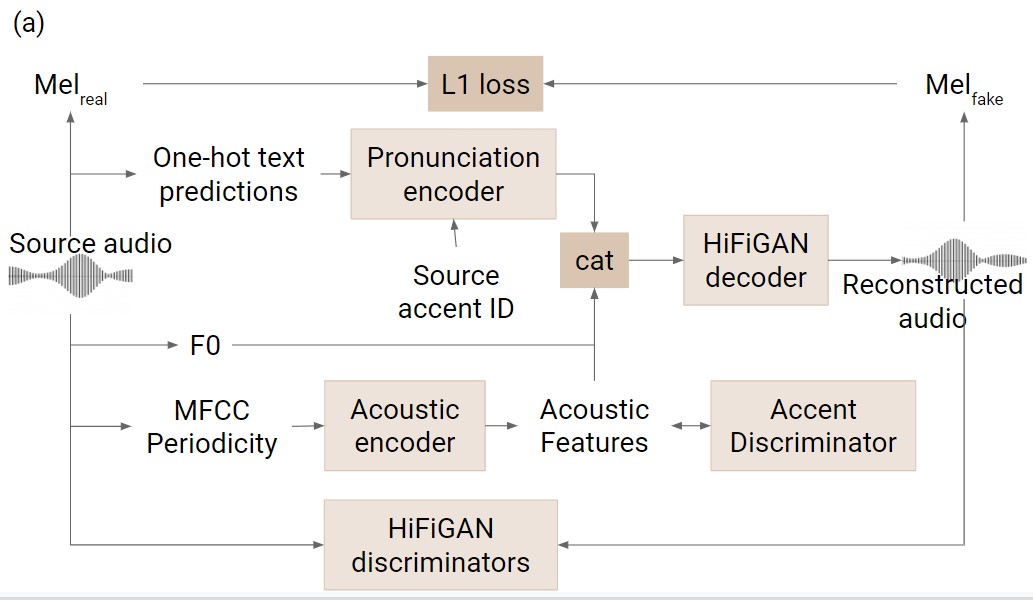}}
\end{minipage}
\begin{minipage}[b]{1.0\linewidth}
  \centering
  \centerline{\includegraphics[width=8.5cm]{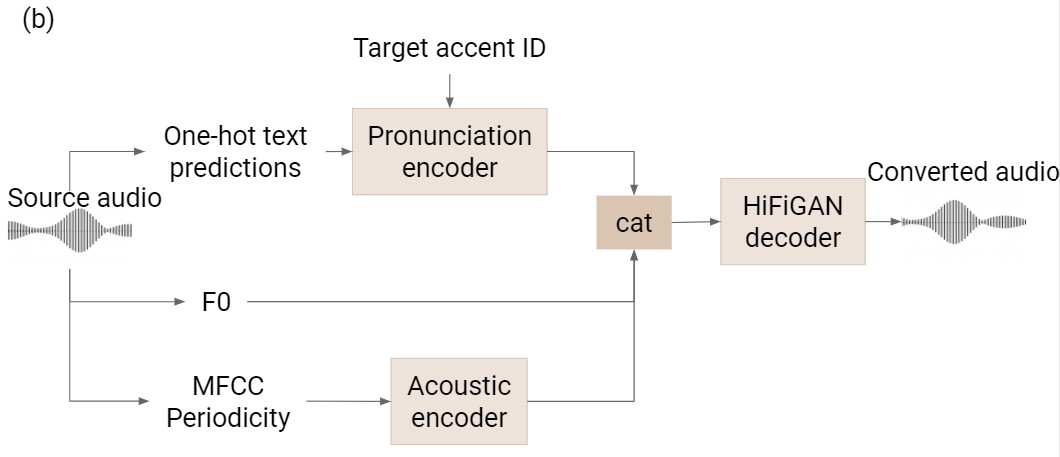}}
\end{minipage}
\caption{(a) Training (b) During inference, we convert accent by feeding in the target accent ID to the pronunciation encoder.} 

\label{fig:model_architecture}

\end{figure}

Our proposed system during training and inference is shown in Fig. \ref{fig:model_architecture}. Unlike \cite{tianjin_mandarin} and \cite{key_baseline}, we do not train an accented ASR model with text labels corresponding to accented speech. Instead, we use an off-the-shelf wav2vec2.0 checkpoint\footnote{Specifically, we downloaded the checkpoint corresponding to \text{"Wav2Vec 2.0 Large (LV-60 + CV + SWBD + FSH)"} from the link \text{https://github.com/facebookresearch/fairseq/tree/main/examples/wav2vec.}} that has been pre-trained using self-supervised learning and fine-tuned on ASR task using $960$ hours of LibriSpeech data\cite{wav2vec2}. 

Let $\mathbf{x}$ denote the input audio waveform, $\mathbf{\hat{x}}$ the output audio, and $M$ the function that transforms a waveform into the corresponding 80-dimensional mel-spectrogram. We train our model to minimize the reconstruction loss in (\ref{eq:reconstruction_loss})
\begin{equation}\label{eq:reconstruction_loss}
  L_{mel} = \mathbb{E}_{\mathbf{x} \in \mathcal{X}}[||M(\mathbf{x}) - M(\hat{\mathbf{x}})||_1]  
\end{equation}

\subsection{Pronunciation Encoder}
\label{ssec:content_encoder}
\begin{figure}[htb]

\begin{minipage}[b]{1.0\linewidth}
  \centering
  \centerline{\includegraphics[width=8.5cm]{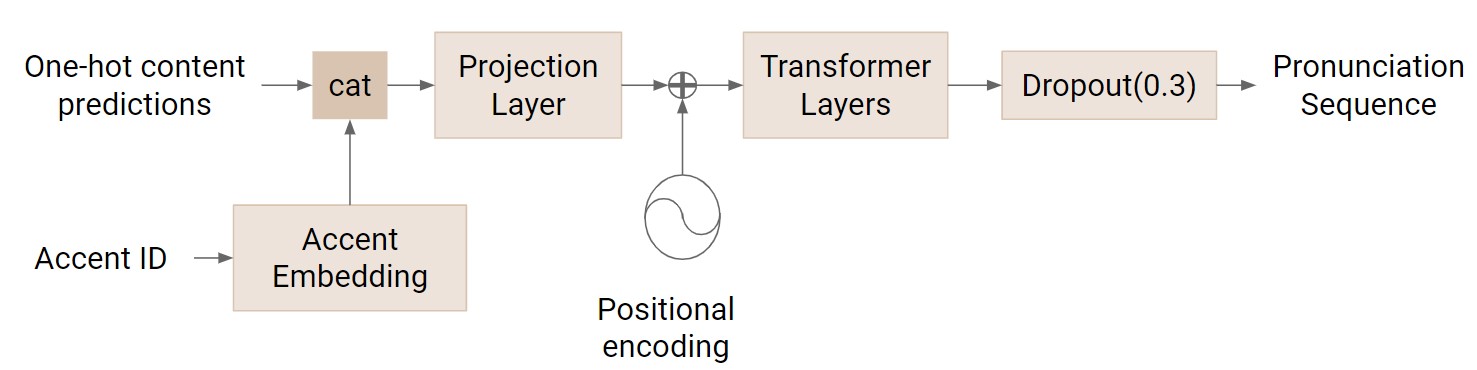}}
\end{minipage}
\caption{The pronunciation encoder.}
\label{fig:pro_encoder}
\end{figure}

Inspired by many early works that have used phonetic posteriograms (PPG) to capture accent-dependent features, we use a pronunciation encoder, as shown in Fig. \ref{fig:pro_encoder}, to synthesize accent-dependent pronunciation sequence given text predictions and accent ID \cite{PPG_ac, PPG_ac2}. For each accent ID, the pronunciation encoder learns a unique embedding, which is concatenated with every frame of character-level wav2vec2.0 prediction. The four transformer layers with 8-head attention mechanism accounts for how context affects pronunciations. Applying a dropout of $0.3$ promotes the decoder to rely more on the acoustic encoder for non-accent related voice features.

\subsection{Acoustic Encoder}
\label{ssec:acoustic_encoder}
The acoustic encoder maps the mel-frequency cepstrum coefficients (MFCC) and periodicity features to a single 256-dimensional vector. The acoustic encoder consists of four convolution layers with kernel sizes (5, 3, 3, 1) and dilations (1, 2, 1, 1), a self-attention layer, and finally an average pooling layer over time to output a single vector. Instead of using an absolute positional embedding, we use the convolution layers to act as relative positional embedding for the self-attention layer as in \cite{wav2vec2}. 

We use adversarial training to remove accent information from the output of the acoustic encoder. All the accents are labeled as either native or foreign. The accent discriminator, consisting of two fully connected layers, learns to predict 1 if the source audio accent is native and 0 if foreign, while the acoustic encoder tries to force the accent discriminator to predict 1 all of the time. Mathematically, $\mathcal{N}$ the set of audios with native English accent, $\mathcal{F}$ the set with foreign accents, and $\mathcal{X} = \mathcal{N} \cup \mathcal{F}$ the entire set of training data. Let $\mathbf{z}$ be the output of the acoustic encoder. The accent discriminator (AD) attempts to minimize $L_{AD}$ from (\ref{eq:accend_clf_loss}) while the acoustic encoder tries to minimize $L_{AD,adv}$ from (\ref{eq:accend_adv}). 

\begin{equation}\label{eq:accend_clf_loss}
    L_{AD} = - \mathbb{E}_{\mathcal{N}}[\log(AD(\mathbf{z}))] - \mathbb{E}_{\mathcal{F}}[\log(1 - AD(\mathbf{z}))]
\end{equation}
\begin{equation}\label{eq:accend_adv}
    L_{AD,adv} = - \mathbb{E}_{\mathcal{F}}[\log(AD(\mathbf{z}))]
\end{equation}


\subsection{HiFiGAN-based Voice Decoder}
\label{ssec:hifigan_decoder}
Finally, we re-combine the accent-dependent pronunciation encodings, the acoustic features, and F0 sequence. We use a modified HiFiGAN to invert the processed features back to audio waveform \cite{hifigan}. Our modifications to the HiFiGAN architecture include adding an additional convolution layer with kernel size 11, modifying the number of input channels to HiFiGAN, and modifying the upsampling rates so that the output length matches the source audio length. 

We use the same multi-scale discriminator (MSD) and multi-period discriminator (MPD) as in \cite{hifigan} to encourage the synthesis of natural sounding audio. Mathematically, if we consider the HiFiGAN discriminators, MSD and MPD, as one discriminator $HD$, the HiFiGAN discriminators try to minimize $L_{HD}$ in (\ref{eq:hifigan_disc_loss}).
\begin{equation}\label{eq:hifigan_disc_loss}
  L_{HD} = \mathbb{E}_{\mathcal{X}} [(HD(\mathbf{x}) - 1)^2 + (HD(\hat{\mathbf{x}}))^2]  
\end{equation}

As in \cite{hifigan}, the adversarial loss $L_{HD,adv}$, and the feature mapping loss functions, $L_{FM}$ applied to the rest of the model in (\ref{eq:hifigan_adv_loss}) and (\ref{eq:hifigan_fm_loss}).
\begin{equation}\label{eq:hifigan_adv_loss}
  L_{HD,adv} = \mathbb{E}_{\mathcal{X}}(HD(\hat{\mathbf{x}}) - 1)^2]  
\end{equation}
\begin{equation}\label{eq:hifigan_fm_loss}
  L_{FM} = \mathbb{E}_{\mathcal{X}} [\sum_{i=1}^T \frac{1}{N_i} ||HD^i(\mathbf{x}) - HD^i(\hat{\mathbf{x}})||]  
\end{equation}

where $HD^i$, $N_i$ denote the features and the number of features in the $i^{th}$ layer of the HiFiGAN discriminators.

\section{Experiments}
\label{sec:experiments}
\subsection{Data Sets}
Our training data is summarized in Table \ref{table:datasets}. Our training data set includes $8$ different accents: American (AM), Arabic (AR), British (BR), Hindi (HI), Korean (KO), Mandarin (MA), Spanish (SP), and Vietnamese (VI). We consider AM and BR accents as native and the rest foreign. Since our data set is highly unbalanced across different accents, we assign different weights to each subset. A weight of $n$ assigned to a subset means that audio clips from that subset will appear around $n$ times during one epoch of training. The weights can also be found in Table \ref{table:datasets}. 

\begin{table*}[t]
  \caption{Training Data Descriptions \cite{libritts, vctk_dataset, speech_accent_archive, l2arctic, indic_tts}}\label{table:datasets}
  \centering
    \begin{tabular}{|p{0.1\linewidth}|p{0.1\linewidth}|p{0.1\linewidth}|p{0.2\linewidth}|p{0.2\linewidth}|p{0.1\linewidth}|}\hline
    Data Set &\makebox{Accents}&\makebox{Duration(hrs)}&\makebox{Speakers} &\makebox{Prompts}&\makebox{Weight}\\\hline\hline
    LibriTTS &AM&$585$&$2456$& LibriSpeech text &1\\\hline
    VCTK & BR &$42$&$109$& Newspaper clippings&6\\\hline
    SAA & BR &$3.7$ &$579$& ``Please call Stella...''&10\\\hline
    L2-Arctic & AR, HI, KO, MA, SP, VI &$24$&$24$ ($2$ male and $2$ female per accent) &ARCTIC prompts&15\\\hline
    Indic TTS &HI&$20.06$&$2$ ($1$ male and $1$ female)& ARCTIC prompts, Fairy tales &2\\\hline
    \end{tabular}
\end{table*}

\subsection{Feature Processing}
All audios are re-sampled at 16kHz and divided into $1.12$s segments for training. We use the YAAPT algorithm to extract the F0 sequence at a frame shift of $5ms$ and window length of $20ms$ \cite{f0_extract}. We up-sampled the outputs of the pronunciation encoder by repeating each time frame $4$ times since text predictions from wav2vec2.0 are extracted at a $20ms$ frame shift. 

\subsection{Training Configurations}
We trained our model with learning rate=$0.0002$ and decayed our learning rate by factor of $0.999$ every $1000$ iterations. We used the AdamW optimizer with $\beta_1=0.8$ and $\beta_2 = 0.99$. For adversarial training, we optimized the accent discriminator for the first $50000$ iterations before applying the adversarial loss to the acoustic encoder. The model was trained for a total of $3$ million iterations using batch size $16$. 

\subsection{Evaluation Set-up}
We conducted mean opinion score (MOS) tests to evaluate the performance of the proposed model on audio naturalness and speaker similarity and an XAB test to evaluate the results of accent conversion. A different set of $100$ listeners participated in each listening test, and we recruited an additional $24$ ``experts" to evaluate accent conversion results. As in \cite{key_baseline}, all audios were shuffled before being presented to the listeners. 

We trained an ablation model by disabling the accent discriminator. We used the same 20 prompt examples from the VCTK dataset used by \cite{key_baseline} for evaluation. In our case, instead of using all examples from the Hindi accented female speaker \textit{p248}, we replaced the first 10 examples with the corresponding ones from a male British accented speaker \textit{p226}. We excluded audios from p248 and p226 from our training set so that they are unseen speakers. Using our proposed model (P) and the ablation model (AB), we converted the $20$ audio clips to American (AM), Hindi (HI), and Korean (KO) accents.

\section{Results}
\label{sec:results}
\subsection{Audio Quality}
The listeners were asked to rate the audio quality for the original (O) and the converted (AB, P) clips on a five-point scale (1-bad, 5-excellent). As seen in Table \ref{table:mos_results}, the proposed model maintains a comparable quality to the original audio. We hypothesize it does better than the ablation model because accent invariance in the acoustic encoder improves generalization.

\subsection{Speaker Similarity}
The raters first listen to a reference audio from each speaker with different text content. For each audio clip, the listeners were asked to rate how close did the voice sound to that of the reference audio's on a five-point scale. The listeners were instructed to disregard the accent and recording conditions. Table \ref{table:mos_results} demonstrates that most listeners believed that the voice in the audios converted using the models sound almost as similar to the reference audio as other original audios from the same speaker.

%
%

\begin{table}
  \caption{MOS Study Results (with $95\%$ confidence interval)}\label{table:mos_results}
  \centering
    \begin{tabular}{|c|c|c|}\hline
    Samples &\makebox{Audio Quality}&\makebox{Speaker Similarity}\\\hline\hline
    Original &$3.87 \pm 0.07$ & $4.28 \pm 0.06$\\\hline
    Ablation & $3.27 \pm 0.12$ & $3.89 \pm 0.10$\\\hline
    Proposed & $3.62 \pm 0.09$ & $4.05 \pm 0.09$\\\hline
    \end{tabular}
\end{table}

\subsection{Accent Conversion}
\begin{figure}[htp]
\begin{minipage}[b]{1.0\linewidth}
  \centering
  \centerline{\includegraphics[width=8.5cm]{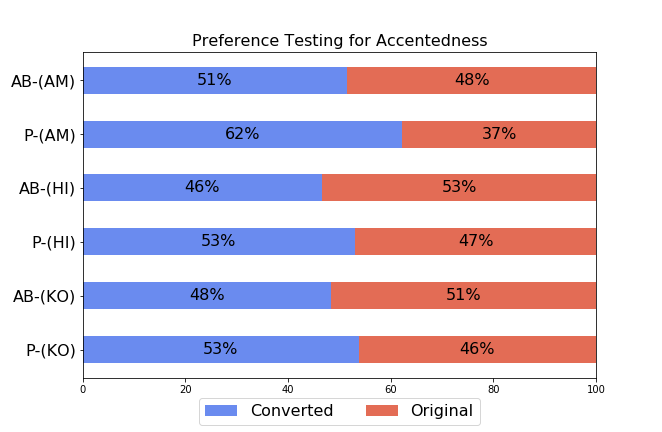}}
  \medskip
\end{minipage}

\caption{Preference results from $100$ random listeners. Inconclusiveness  on accents other than American suggests that recognizing attributes of a specific accent may need more exposure than obtained from listening to a couple reference clips. }
\label{fig:xab_result}
\end{figure}

\begin{figure}[htp]

\begin{minipage}[b]{1.0\linewidth}
  \centering
  \centerline{\includegraphics[width=8.5cm]{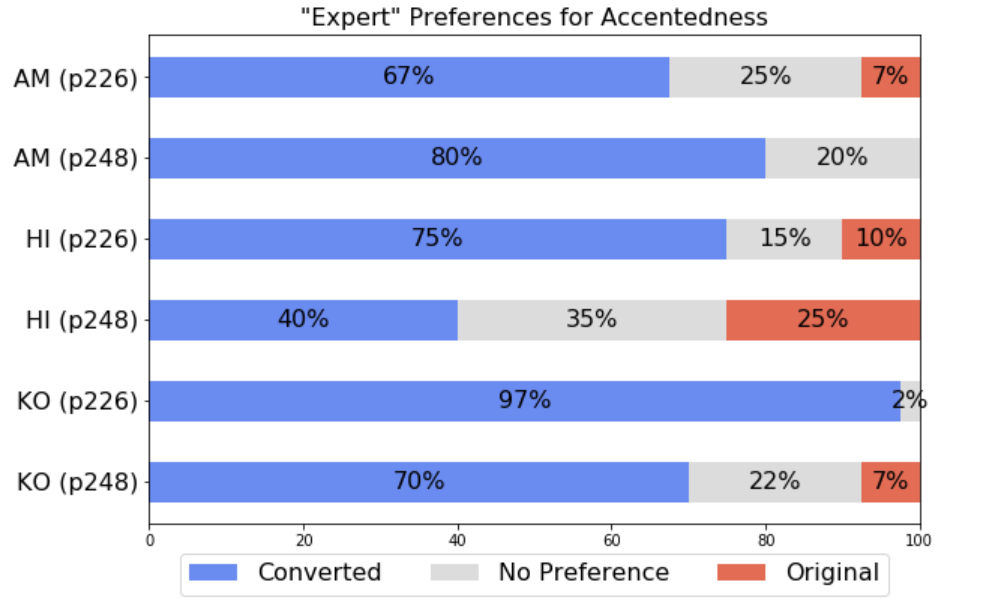}}
\end{minipage}
\caption{Preference results from raters having substantial exposure to respective accents (4 raters per accent). With the exception of HI-to-HI conversion for p248, the listeners strongly preferred the audios converted by our proposed model. Preference is stronger when converting from native to non-native accents or vice versa.}
\label{fig:xab_result_expert}
\end{figure}

The listeners first listened to some reference audio clips with the target accent. Then, for each pair containing an original audio clip and its converted audio clip, the listeners were asked to choose the one that has a closer accent to the reference audio. As seen from the results in Fig. \ref{fig:xab_result}, for every target accent, the proposed model performs better than the ablation model. Significantly more listeners preferred the converted audios as sounding more American than the original audios. However, the listeners seemed unsure about whether the converted audios sounded more Korean or Hindi, likely because it is difficult to identify an accent after listening to only a few audio clips. The goal of conversion to other foreign accents is to help a non-native listener understand someone with a different foreign accent or to create a better entertainment value for the people knowledgeable of the desired accent. Therefore, we recruited $24$ additional ``expert" participants who were born and raised in USA, India, and Korea, to judge the corresponding accents. The results shown in Fig. \ref{fig:xab_result_expert} indicate that the listeners who are very familiar with the target accent were confident that the proposed model converted audios to sound like the target accent. 

\section{Conclusion}
\label{sec:discussion}
In this paper, we presented a novel accent conversion model that converts an unseen speaker's utterance with an arbitrary accent to utterances with multiple different target accents\footnote{ Demo samples at \url{https://accent-conversion.github.io}}. The model is able to noticeably convert the accent of input audios while preserving speaker identity and audio quality. Future work may improve the perceptive accent accuracy of the converted audio by predicting pitch based on the accent.

\bibliographystyle{IEEEbib}
\bibliography{references}

\end{document}